\documentclass[apj]{emulateapj}

\shorttitle{Intergalactic Gas in Groups of Galaxies}
\slugcomment{Accepted for publication in The Astrophysical Journal}
\shortauthors{Freeland \& Wilcots}

\def\mr{\mathrm}

\def\kms{\mr{~km\ s}^{-1} }

\def\hi{H{\sc i}~}
\def\himf{H{\sc i}MF~}

\def\msol{\mr{M}_{\odot}}

\def\kmsmpc{\mr{\ km}\ \mr{s}^{-1}\ \mr{Mpc}^{-1}}

\def\cmc{\ \mr{cm^{-3}}}

\def\gcmc{\ \mr{g \cmc}}
\def\mjbm{\mr{~mJy~beam}^{-1}} 
\begin{document}
\title{Intergalactic Gas in Groups of Galaxies: \\Implications for Dwarf Spheroidal Formation and The Missing Baryons Problem}
\author{E. Freeland}
\affil{George P. and Cynthia W. Mitchell Institute for Fundamental Physics and Astronomy, Dept. of Physics and Astronomy, Texas A\&M University,College Station, TX 77843}
\author{E. Wilcots}
\affil{Department of Astronomy, University of Wisconsin, Madison, WI 53706}
\email{freeland@physics.tamu.edu, ewilcots@astro.wisc.edu}

\begin{abstract}

Radio galaxies with bent jets are predominantly located in groups and clusters of galaxies.  We use bent-double radio sources, under the assumption that their jets are bent by ram-pressure, to probe intragroup medium gas densities in galaxy groups.  This method provides a direct measurement of the intergalactic gas density and allows us to probe intergalactic gas at large radii and in systems whose intragroup medium (IGM) is too cool to be detected by the current generation of X-ray telescopes.  We find gas with densities of $10^{-3}-10^{-4}\cmc$ at group radii from $15-700$ kpc.  A rough estimate of the total baryonic mass in intergalactic gas is consistent with the missing baryons being located in the intragroup medium of galaxy groups.  The neutral gas will be easily stripped from dwarf galaxies with total masses of $10^{6}-10^{7}\ \msol$ in the groups studied here.  Indications are that intragroup gas densities in less-massive systems like the Local Group should be high enough to strip gas from dwarfs like Leo T and, in combination with tides, produce dwarf spheroidals.  
\end{abstract}


\section{Introduction}

Galaxy groups are ubiquitous, intermediate density structures, spanning the range between isolated field galaxies and rich clusters \citep{1983ApJS...52...61G,1987ApJ...321..280T,2004MNRAS.348..866E,2010A&A...514A.102T}.   According to the hierarchical scenario of the formation of large-scale structure, groups are the building blocks of rich clusters of galaxies \citep{2008A&A...489...11B,2009MNRAS.400..937M}.  Thus, understanding the physical properties of clusters and their member galaxies requires knowledge of the extent to which galaxy evolution occurs in the group environment.  The effectiveness of physical processes that may be responsible for this evolution, like ram-pressure stripping and strangulation, is a function of the density of intergalactic gas.  There are only a handful of ways to detect this gas, namely X-ray observations which are limited to the more massive groups and UV absorption line studies which probe gas with a limited temperature range.  We discuss here a complementary method, applicable in even low mass groups, which can determine the intergalactic gas density using bent-double radio sources.

Groups are likely to contain a significant fraction of the baryons in the local universe \citep{1998ApJ...503..518F,2000ARA&A..38..289M}.  Attempts to reconcile the baryon content of the local universe with the baryon density seen at high redshift have failed to locate the majority of baryons \citep{2004ApJ...616..643F}.  Specifically, the baryon deficit appears to scale with potential well depth such that the baryon content of massive clusters is nearly as expected while groups and individual galaxies are lacking \citep{2003ApJ...585L.117B, 2010ApJ...719..119D}.  These locally ``missing baryons'' are predicted by simulations to exist in a warm-hot intergalactic medium (WHIM) that pervades large-scale structures \citep{1999ApJ...514....1C,2001ApJ...552..473D}.  In simulations this gas is shock heated during structure formation, leading to temperatures in the range $10^5 < T < 10^7$ K and a broad density distribution that is peaked at $10-20$ times the critical density of the universe \citep{2006ApJ...650..560C}.  Ultraviolet (UV) and X-ray absorption line detections certainly confirm the existence of the highly ionized WHIM (see the review by \citet{2007ARA&A..45..221B}) but the spatial distribution of this gas inside and outside of galaxies is still unconstrained.  

We explore the potential impact of the presence of a widespread intergalactic medium with with densities $\sim 10^{-3}-10^{-4} \cmc$ by looking at the fate of the gas content of dwarf spheroidal galaxies.  Dwarf spheroidal (dSph) galaxies are dark-matter dominated, gas poor, low surface brightness, and have stellar populations with old to intermediate ages.  Evidence for the role of environment in the evolution of dSphs is seen in the morphology-density relations which exist for the Local Group and nearby groups \citep{1994AJ....107.1328V,2009AJ....138.1037C,2003AJ....125..593S,2009AJ....137.3038B}; dSphs are preferentially found near large galaxies.  Their progenitors are thought to be similar to dwarf irregular (dIrr) galaxies which were processed by their environment, causing angular moment loss and gas stripping \citep{2003AJ....125.1926G,2010AdAst2010E..25M}.  This is supported by the observation that structurally, when comparing surface brightness profile shapes, dSphs, dIrrs, and galaxy disks form a family that is distinct from that of elliptical and dwarf elliptical galaxies \citep{1983ApJ...266L..21L,1985ApJ...295...73K,1987nngp.proc..163K,2009ApJS..182..216K}.  However, dSphs have slightly higher metallicities and surface brightnesses than current dIrrs, indicating that they have undergone comparatively more early and rapid star formation episodes \citep{1998ARA&A..36..435M}.  A third class of dwarf, known as ``transition-type'', possesses a mixture of dIrr and dSph properties and is likely to be the progenitor of future dSph galaxies.  Interestingly, these transition dwarfs also show a morphology-density relation in the Centaurus A group, with average distances from large galaxies intermediate between those of dSphs and dIrrs \citep{2009AJ....138.1037C}.  Ram pressure stripping, which depends on the density of the intergalactic gas in these groups, is the favored mechanism for the removal of the interstellar neutral gas from dSph progenitors \citep{2003AJ....125.1926G,2010AdAst2010E..25M}.

Radio galaxies with bent jets are preferentially located in groups and clusters of galaxies, whereas straight and single component radio sources are not \citep{2011AJ....141...88W}.  We present measurements of intergalactic gas densities in groups of galaxies, assuming that the jets of these radio galaxies are bent backward by ram-pressure, for five new sources.  These are groups for which no X-ray data exist or which are undetected in current observations.  We include the two radio galaxies from \citet{2008ApJ...685..858F} (hereafter F08) as we have slightly altered our original set of assumptions about the bulk flow speeds along the radio galaxy jets.  The original results change only marginally whereas the derived densities are now more constrained.  Sources have been renamed, as follows, to allow easy referencing:   CGCG 156-060 (S3), SDSS J154849.35+361035.3 (S4), SDSS J112038.52+291234.1 (S5), SDSS J115434.81+363539.8 (S6), SDSS J212616.07-071046.3 (S7) and FIRST J124942.2+303838 (S1), SDSS J085331.86+23400.0 (S2) from F08.   Table \ref{tab:sour} includes source coordinates and properties as well as results.

We use a Hubble constant of $73 \kmsmpc$ and velocities which are corrected with respect to the Cosmic Microwave Background when determining distances.  This Hubble constant corresponds to $\rho_{crit}=3c^2H_0^2/8\pi G=1.0 \times 10^{-29} \gcmc$.

\section{Observations} 

\subsection{Radio Data} 

We chose bent-double radio galaxies for this analysis primarily from two
studies that utilize the VLA FIRST ((Faint Images of the Radio Sky at
Twenty-Centimeters) survey \citep{1995ApJ...450..559B}: S4, S6 and S7 are from
a sample devised by pattern-recognition \citep{2006ApJS..165...95P} while S1,
S2 and S5 are from a sample chosen by visual inspection \citep[][hereafter
B01]{2001AJ....121.2915B}.  The nearest source, S3, was first brought to our
attention in \citet{2005ApJ...626..748Y} and is not in either of the above
studies because the FIRST data resolve out most of its extended jet structure.
When culling our sample from the above two studies we placed a high priority on
sources with symmetrical jet structure and brightness with an eye toward
minimizing projection effects.  We then chose only sources that were not
obvious cluster members by visually inspecting their environment using optical
sky surveys or which were characterized by B01 as existing in groups or the
field.   

For all sources, except S3, we use FIRST radio continuum data at 1420 MHz to
measure the jet width, radius of curvature, and determine the internal
synchrotron pressure in each radio source.  The data have a resolution of $5.4
\arcsec \times 5.4 \arcsec$ and a typical rms of $0.15 \mjbm$. 

Data for S3 were procured from the Very Large Array (VLA) archive from program
ID AR402.  This source was observed for half an hour in the B configuration in
February of 2000.  The final map has a resolution of $4.8 \arcsec \times 4.6
\arcsec$ and an rms noise of $0.2 \mjbm$.


\subsection{Optical Observations} 

Multi-object spectroscopy was performed using HYDRA on the WIYN 3.5m telescope
on Kitt Peak.  Fibers from the blue cables were placed on galaxies chosen from
Sloan Digital Sky Survey (SDSS) photometric redshifts to have a redshift
similar to the radio source while additional fibers were placed on blank sky
positions.  The 600@10 Zepf grating was used in first order, giving spectra
with a dispersion of $1.4$ \AA\ per pixel covering a wavelength range from 4600
to 7200 \AA.  CuAr lamp calibration spectra were used for the wavelength
calibration and an averaged sky spectrum was subtracted from each object
spectrum.   The data were reduced in IRAF using the {\it dohydra} package.
Cross correlation of object spectra with galaxy template spectra was performed
using the RVSAO \citep{1998PASP..110..934K} package in IRAF.  

We combined redshift information from our WIYN spectroscopy with spectroscopic
redshifts from the Sloan Digital Sky Survey DR6 and DR7 data releases
\citep{2008ApJS..175..297A} and NASA Extragalactic Database (NED).  For the S1,
S4, S5, S7 groups about half of the redshifts are from WIYN and the other half
are from SDSS or NED.  For the S3 and S6 groups all redshifts are from SDSS.
The only photometric redshift used for the new sources presented here is that
of radio source S7.

\begin{figure*}
\plotone{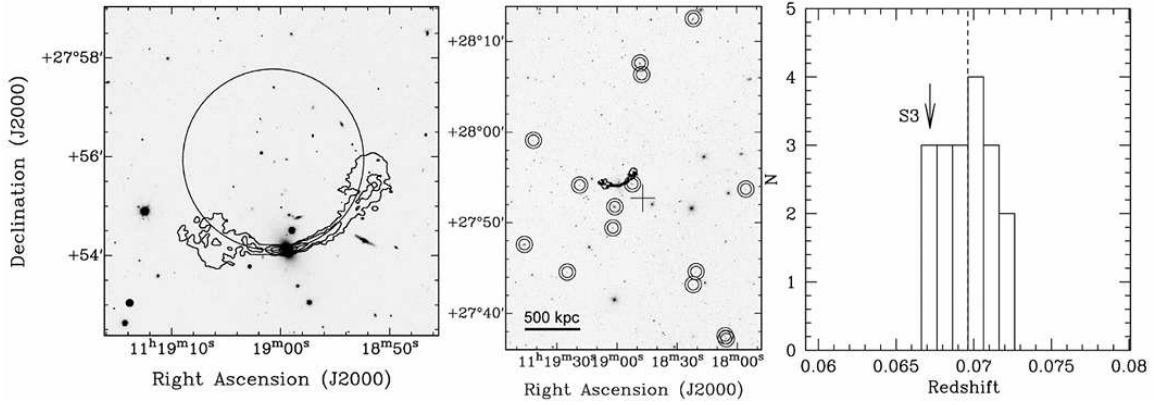}
\caption{VLA 1420 MHz radio continuum contours overlaid on an SDSS r band optical image of CGCG 156-060 (S3).  The lowest contour is $0.5 \mjbm$ and contours increase by factors of 2.5, the beam is shown in the lower left corner. In the panel on the left the radius of curvature is illustrated.  In the central panel, neighboring galaxies with spectroscopic redshifts are circled twice, two are beyond the range of the image.  The cross indicates the average position of the group center.  In the panel on the right, the redshift histogram is shown, the dashed line indicates the group redshift and the arrow the redshift of the bent-double radio source.   Redshift bins are $300 \kms$ wide.  The physical scale at the distance of this group is $80$ kpc $\mathrm{arcmin}^{-1}$.  We measure an IGM density of $2\pm 1 \times 10^{-4} \cmc$ ($20 \rho_{crit}$) at the location of S3.}  \label{fig:cgcg}
\end{figure*}

\begin{figure*} 
\plotone{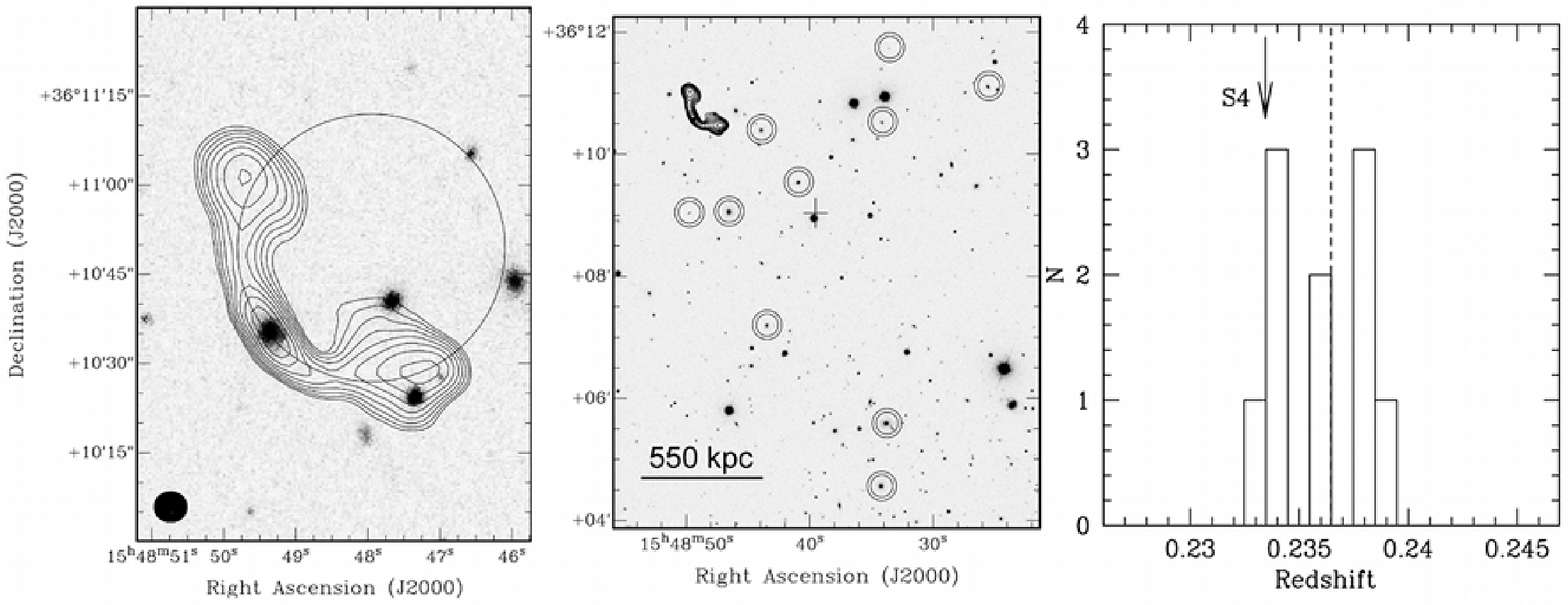}
 \caption{GMRT 610 MHz radio continuum contours overlaid on an SDSS i band optical image of SDSS J$154849.35+361035.3$ (S4).  The lowest contour is $1 \mjbm$ and increases by $\sqrt{2}$, the beam is shown in the lower left.  In the panel on the left the radius of curvature is illustrated.  In the central panel, neighboring galaxies with spectroscopic redshifts are circled twice.  The cross indicates the averaged position of the group center.  In the panel on the right, the redshift histogram for the circled galaxies is shown.  The dashed line indicates the group redshift and the arrow the redshift of the bent-double radio source.  Redshift bins are $300 \kms$ wide.  The physical scale at the distance of this group is $280$ kpc $\mathrm{arcmin}^{-1}$.  We measure an IGM density of $5\pm 2 \times 10^{-4}\cmc$ ($50 \rho_{crit}$) at the location of S4.}  \label{fig:606}
\end{figure*}

\begin{figure*}
\plotone{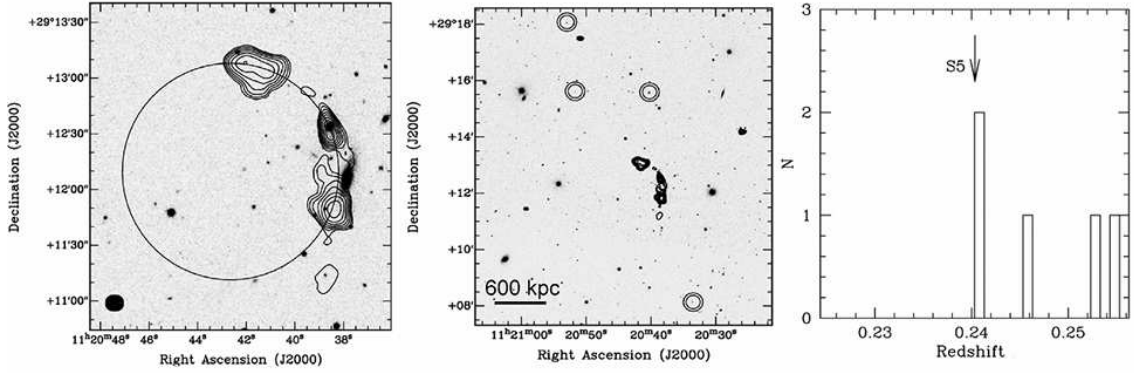}
 \caption{GMRT 610 MHz radio continuum contours overlaid on an SDSS i band optical image of SDSS J112038.52+291234.1 (S5).  The lowest contour is $0.7 \mjbm$ and increases by $\sqrt{2}$, the beam is shown in the lower left.  In the panel on the left the radius of curvature is illustrated.  In the central panel, neighboring galaxies with spectroscopic redshifts are circled twice.  In the panel on the right, the redshift histogram for the circled galaxies is shown, the arrow indicates the redshift of the bent-double radio source.  Redshift bins are $300 \kms$ wide.  The physical scale at the distance of this group is $290$ kpc $\mathrm{arcmin}^{-1}$.  As we are unable to constrain the velocity of S5 we give the IGM density near the bent-double source as a function of its velocity in $\kms$,  $(23 \pm 7) / v^2$. }  \label{fig:195}
\end{figure*}

\begin{figure*} 
\plotone{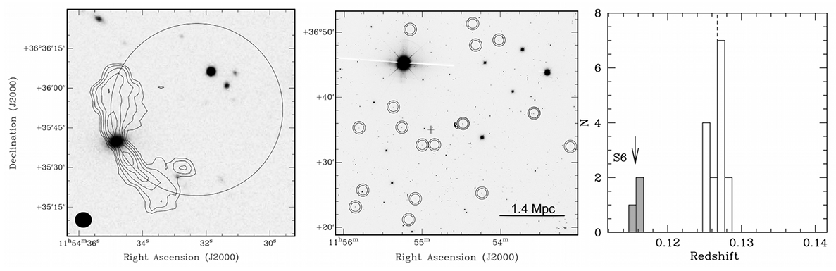}
 \caption{FIRST 1420 MHz radio continuum contours overlaid on an SDSS i band optical image of SDSS J115434.81+363539.8 (S6). The lowest contour is $0.45 \mjbm$ and increases by $\sqrt{2}$, the beam is shown in the lower left corner.  In the panel on the left the radius of curvature is illustrated.  In the central panel, neighboring galaxies with spectroscopic redshifts are circled twice.  The cross indicates the averaged position of the group center.  In the panel on the right, the redshift histogram for the circled galaxies is shown.  The dashed line indicates the nearby group redshift and the arrow the redshift of the bent-double radio source, redshift bins are $300 \kms$ wide.  The three sources in the grey histogram are ringed in grey in the central panel.  The physical scale at the distance of this group is $140$ kpc $\mathrm{arcmin}^{-1}$.  As we are unable to constrain the velocity of S6 we give the IGM density near the bent-double source as a function of its velocity in $\kms$,  $(52 \pm 16) / v^2$}  \label{fig:336}
\end{figure*}

\begin{figure*} 
\plotone{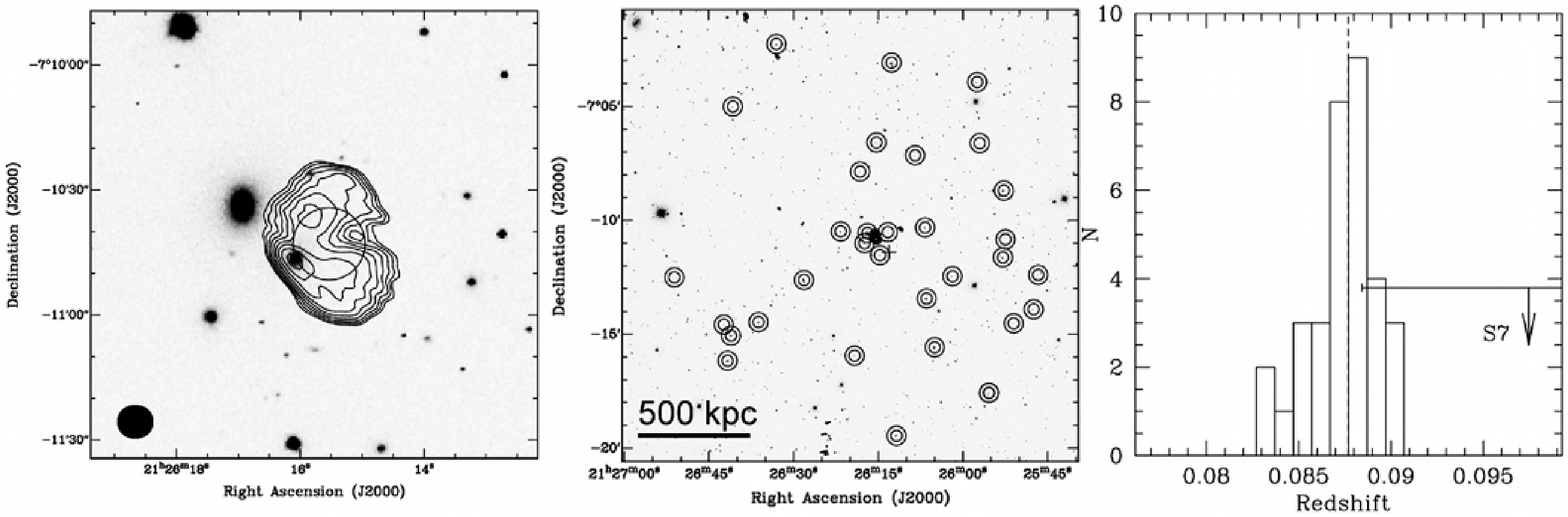}
 \caption{FIRST 1420 MHz radio continuum contours overlaid on an SDSS i band optical image of SDSS J212616.07-071046.3 (S7). The lowest contour is $1 \mjbm$ and increases by $\sqrt{2}$, with the beam shown in the lower left.  In the panel on the left the radius of curvature is illustrated.  In the central panel, neighboring galaxies with spectroscopic redshifts are circled twice.  The cross indicates the averaged position of the group center.  In the panel on the right, the redshift histogram for the circled galaxies is shown centered on the redshift of the radio source.  The dashed line indicates the group redshift and the arrow the photometric redshift, with errors, of the bent-double radio source.  Redshift bins are $300 \kms$ wide.  The physical scale at the distance of this group is $105$ kpc $\mathrm{arcmin}^{-1}$.  We measure an IGM density of $2\pm 1 \times 10^{-3}\cmc$ ($200 \rho_{crit}$) at the location of S7. }  \label{fig:700}
\end{figure*}

\begin{deluxetable*}{lcccccccc}
\tablecaption{Source Information\label{tab:sour}}
\tabletypesize{\scriptsize}
\tablehead{
 & \colhead{z} & \colhead{$R_{group}\tablenotemark{a}$} & \colhead{$L_{1440}\tablenotemark{b}$} & \colhead{$h$} & \colhead{$R_{bend}$} & \colhead{$v_{gal}$} & \colhead{$P_{min,jet}$} & \colhead{$n_{\mbox{\tiny IGM}}\tablenotemark{c}$} \\
 &             & (kpc)                 & ($\mathrm{W\ Hz}^{-1}$) & (arcsec) & (arcsec) & ($\kms$) &  ($10^{-11}\ \mathrm{dynes~cm}^{-2}$) & ($ \cmc$) }  
\startdata
S1   
& 0.194    & 300	& $1.6 \times 10^{25}$  & 7.0 $\pm$ 0.3  & $13 \pm 2$   & $250_{-20}^{+100}$  & 0.9 $\pm$ 0.2 & $3 \pm 2 \times 10^{-3}$	\\  

S2   
& 0.306    & 2000	& $1.88 \times 10^{25}$ & 6.5 $\pm$  1   & $23 \pm 2$   & $570 \pm 60$        & 0.6 $\pm$ 0.2 & $5 \pm 4 \times 10^{-4}$       	\\ 

S3   
& 0.067    & 270	& $3 \times 10^{23}$    & 8.0 $\pm$0.5   & $110 \pm 15$ & $430_{-46}^{+63}$   & 1.4 $\pm$ 0.6 & $2 \pm 1 \times 10^{-4}$ 	\\  

S4   
& 0.233475 & 700	& $9.5 \times 10^{24}$  & 6.0 $\pm$0.2   & $24 \pm 2$   & $550^{+120}_{-80}$  & 1.7 $\pm$ 0.3 & $5 \pm 2  \times 10^{-4}$	\\  

S5   
& 0.240341 & -		& $6.4 \times 10^{24}$  & 10 $\pm$ 1     & 58 $\pm$ 3   & unconstrained       & 0.4 $\pm$ 0.1 & $(23 \pm 7)/v^2$		 \\  

S6   
& 0.115606 & -		& $9 \times 10^{23} $   & 9.5$\pm$ 1.5   & $33\pm2$     & unconstrained       & 0.6 $\pm$ 0.1 & $(52\pm 16)/v^2 $		 \\  

S7   
&0.097$\pm$0.009& 15	& $2\times10^{24}$      & 12 $\pm$ 2     & 10 $\pm$ 2   & $490^{+100}_{-70}$  & 1.4 $\pm$ 0.3 & $2 \pm 1 \times 10^{-3}$ 	 

\enddata
\tablenotetext{a}{Distance of the bent-double source from the group center.}
\tablenotetext{b}{$L_{1440}$ is calculated from the FIRST data or VLA archival data in the case of S3.}
\tablenotetext{c}{The errors on $n_{\mbox{\tiny IGM}}$ conservatively include a symmetric contribution using the largest of the errors on $v_{gal}$ as there is no generally accepted way to propagate asymmetric errors.  For sources S5 and S6 where the velocity of the radio source is unconstrained this density is given as a function of the radio source velocity in $\kms$.}
\end{deluxetable*}

\section{Method}

In our previous paper, F08, we presented measurements of intergalactic gas densities made using two radio galaxies whose jets are assumed to be bent by ram-pressure as their host galaxy travels through intergalactic gas.  Our method in this paper is the same with one caveat, we no longer assume the jet speeds to be relativistic on kiloparsec scales.  If the bulk flow in the jets was moving relativistically we would expect to see less symmetric radio fluxes in the two jets from beaming effects.  Also, there is increasing evidence that FRI radio source jets are slowed down by entraining material from their surroundings \citep{1984ApJ...286...68B,2002MNRAS.336.1161L,2008MNRAS.386.1709C}.  

For a non-relativistic jet, the Euler equation describes the balance of internal and external
pressure gradients \citep{1979Natur.279..770B,1979ApJ...234..818J,1980AJ.....85..204B},  

\begin{equation} \frac{\rho_{\mbox{\tiny IGM}}
v_{gal}^2}{h}=\frac{\rho_{jet}v_{jet}^2}{R} \end{equation}  where
$\rho_{\mbox{\tiny IGM}} v_{gal}^2$ is the external ram-pressure felt by the
radio galaxy as it travels through the IGM, $\rho_{jet}v_{jet}^2$ is the
pressure inside the jet, $h$ is the width of the jet.  We assume the jets have
cylindrical geometry.  
In the case that the jet width is resolved we deconvolve the true jet size from the
radio beam and the deconvolved width is then used to calculate the source volume when
determining the internal synchrotron pressure as outlined in \citet{1979AJ.....84.1683B}.  

Errors on the jet width for each source represent the range of widths
observed from multiple slices along the length of both jets.  The radius of
curvature, $R$, is found by fitting a circle by eye along the jets through the
core, errors reflect the range of acceptable by-eye fits.  The velocity of the
radio galaxy, $v_{gal}$, is estimated using the velocity dispersion of the
group as $\sqrt{3}\sigma_{v}$. 


We measure the minimum synchrotron pressure, $P_{min}$ in the jets using our
radio data as outlined in \citet{1987ApJ...316...95O}.
\begin{equation}P_{\mr{min}}=(2\pi)^{-\frac{3}{7}}\left( \frac{7}{12} \right)
[c_{12}L_{\mr{rad}}(1+k)(\phi V)^{-1}]^{\frac{4}{7}} \ \ \mr{ergs \ cm^{-3}}
\end{equation} where $c_{12}$ is a constant that depends on the spectral index
and frequency cutoffs \citep{1970ranp.book.....P}.  We take $k=1$ where $k$ is
the ratio of relativistic proton to relativistic electron energy, $\phi=1$
where $\phi$ is the volume filling factor, $V$ the source volume, and
$L_{\mr{rad}}$ the radio luminosity for a given slice through the jet.  The
synchrotron spectrum is integrated from 10 MHz to 10 GHz.  We measure an
internal pressure at multiple positions along each jet, always excluding the
core.   We assume a value of $\alpha = -0.55$ for the spectral index of the
synchrotron emission in the jet \citep{2005ApJ...626..748Y}.  

There is evidence that an additional pressure source may be present from
observations of FRI lobes whose minimum synchrotron pressure does not balance
the observed external pressure as measured by modeling the X-ray emission from
the surrounding IGM
\citep{1984ApJ...286...68B,2002MNRAS.336.1161L,2008MNRAS.386.1709C}.  The
standard assumption is that entrained thermal protons are likely responsible
for the additional energy density above that provided by the relativistic
particles and magnetic fields in the jets.  This entrained material must have a
lower temperature than the surrounding hot gas because the X-ray surface
brightness decreases at the locations of the radio lobes
\citep[e.g.][]{2003MNRAS.346.1041C}.  The additional pressure provided by the
entrained material is not well constrained but it will lead to an increase in
the derived IGM densities.  The IGM densities presented in this paper consider
jet pressures under the minimum energy condition only.  

In this analysis we assume that the radio source is moving in the plane of the
sky and that projection effects do not significantly alter our measurement of
the radius of curvature of the jets.  The measured IGM density depends on the
inverse of the radius of curvature, $R$.  The value of $R$ changes weakly with
the angle of the radio source out of the plane of the sky.  This orientation
can be parametrized by two angles of rotation through axes centered on the
radio galaxy.  We would need to underestimate $R$ by more than a factor of
three to reduce the derived IGM densities and invalidate our conclusions.  An
examination of the projected radius of curvature in the parameter space spanned
by the two rotation angles indicates that orientations which underestimate $R$
by more than a factor of three occur only $7\%$ of the time.  Thus, we are
confident that our conclusions do not depend strongly on the radio source
projection.

With more data projection effects might be examined by comparing the Faraday
rotation signal from the two radio lobes; the lobe closer to the observer would
exhibit less rotation having traveled through less ionized IGM.  Additionally,
much higher resolution radio data of the core could reveal relativistic beaming
effects that, with an accurate assumption of the bulk flow velocity down the
jets, might help constrain the source projection.    

\section{Individual Sources}

The nomenclature for individual sources used here continues a series which began with S1 and S2 in F08.  We include their details here for completeness.  The mapping to longer source names is given at the end of the introduction.  For all sources, values of radius within the group, jet width, radius of curvature of the jets, internal synchrotron pressure, 1440 MHz luminosity, group velocity dispersion, and derived IGM density are given in Table \ref{tab:sour}. 

\subsection{S1}
Originally presented in \citep{2001AJ....121.2915B} and F08 this source is $300$ kpc from the average center of a small group of galaxies whose velocity dispersion is $250_{-20}^{+100} \kms$.  The IGM density at this location is $3 \pm 2 \times 10^{-3} \cmc$.  

\subsection{S2}
Originally presented in \citep{2001AJ....121.2915B} and F08 this source is 2 Mpc from the center of a system of galaxies.  This source is far enough away that we estimate its velocity using the difference in redshift instead of the velocity dispersion of the system which is hard to constrain since most of the redshifts we have are photometric.  With a velocity of $570 \pm 60 \kms$ this sources is probing IGM gas with a density of $5 \pm 4 \times 10^{-4} \cmc$. 

\subsection{S3}
Shown in Figure \ref{fig:cgcg}, this source is located a projected distance of $270$ kpc from the center of a system of galaxies with a velocity dispersion of $430_{-46}^{+63} \kms$.    We calculate an IGM density near the bent-double source of $2 \pm 1 \times 10^{-4} \cmc$.   

\subsection{S4}
This source is on the outskirts of a system of galaxies with a velocity dispersion of $710\pm200 \kms$, as shown in Figure \ref{fig:606}.  If we remove the galaxy with the largest redshift, which is $2\sigma$ from the center of the distribution, then the velocity dispersion is $550^{+120}_{-80} \kms$.  The bent-double radio source lies a projected distance of $700$ kpc from the center of the group.    We calculate an IGM density near the bent-double source of $5 \pm 2 \times 10^{-4} \cmc$.

\subsection{S5}
This source is one of a handful which we have taken from \citet{2001AJ....121.2915B} because of their likelihood of existing in a group of galaxies instead of a cluster.  \citet{2001AJ....121.2915B} use a richness measurement of the local galaxy density, similar to that used for Abell clusters, which is outlined in \citet{1997ApJ...476..489Z}.  This method characterizes the richness of the environment using the number of galaxies within a projected 0.5 Mpc radius of the radio galaxy with absolute magnitude brighter than $M_V = -19$.  
This source, SDSS J112038.52+291234.1, appears to be in an extremely isolated environment (see Figure \ref{fig:195}), although it is likely to be in a pair with nearby SDSS J112051.70+291537.3.  Their velocity difference is only $10 \kms$.  Due to its isolated environment, we are not able to constrain the space velocity of this source.  We give the IGM density near the bent-double source as a function of its velocity in $\kms$,  $(23 \pm 7) / v^2 \cmc$ .

\subsection{S6}
This source, shown in Figure \ref{fig:336}, is slightly offset in redshift from a group at $z=0.127$ whose velocity dispersion is $\sim 300 \kms$.  It has two nearby neighbors in redshift and the velocity dispersion of the radio galaxy and these two neighbors is $130 \kms$.  The velocity difference between the radio galaxy and the center of the nearby system is $3300 \kms$ so we assume that they are not significantly dynamically associated and as a result we are unable to constrain the velocity of the radio source.    We give the IGM density near the bent-double source as a function of its velocity in $\kms$,  $(52 \pm 16) / v^2 \cmc$ .

\subsection{S7}
This source is near the center of a large group whose velocity dispersion is $490^{+100}_{-70} \kms$, see Figure \ref{fig:700}.  We currently do not have a redshift for the radio source although it has an SDSS photometric redshift which is consistent with its membership in the group.  Its projected distance from the average group center is only $0.14 \arcmin$ which corresponds to $15$ kpc.  Due to its projected location near the center of group we do not have a good idea of its actual radial distance from the group center.    We calculate an IGM density near the bent-double source of $2 \pm 1 \times 10^{-3} \cmc$.

\begin{deluxetable}{lccc}
\tablecaption{Generic Dwarf Galaxy Characteristics\label{tab:dwarfs}}
\tabletypesize{\scriptsize}
\tablehead{
Total Mass & \hi Mass & $\sigma_*$ & $n_{gas}$ \\
($\msol$) & ($\msol$)  & ($\kms$)     & ($\cmc$) }
\startdata
$5\times 10^8$  & $5 \times 10^7$  &  25  & 1 \\
$5\times 10^7$  & $5 \times 10^6$  &  15  & 0.5 \\
$5\times 10^6$ &  $5 \times 10^5$  &  10  & 0.5
\enddata
\end{deluxetable}

\begin{figure} 
\plotone{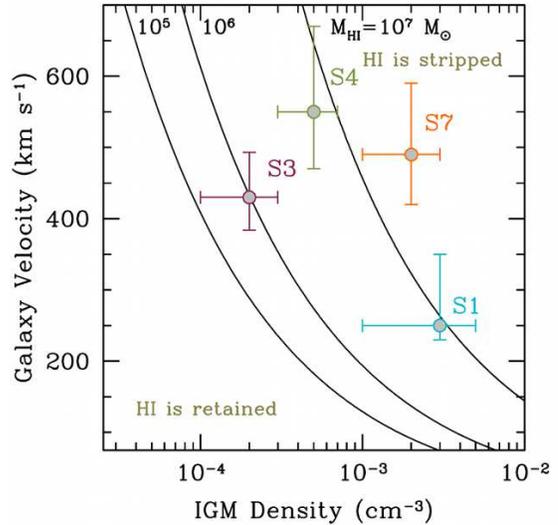}
 \caption{Black lines indicate the combination of velocity and IGM density necessary to instantaneously ram-pressure strip the \hi content of three generic dwarf galaxies with characteristics given in Table \ref{tab:dwarfs} and according to the relation in Equation 3.  In all three cases the total mass of the galaxy is simply $10\times M_{HI}$.  Points indicate the groups studied in this paper.  \textit{Neutral gas is easily stripped from dwarf galaxies with $M_{HI}= 10^{5}-10^{6}$ in the group environment.}} \label{fig:dwarfs}
\end{figure}

\section{Discussion}

In the previous section, and Table \ref{tab:sour}, we have presented measurements of intergalactic gas densities using radio sources located in groups of galaxies.  Assuming that these densities are representative of the IGM in groups in general we now explore the impact this gas can have on gas rich dwarf galaxies, hot gas in galaxy halos, and the baryon fraction.

\subsection{Ram Pressure Stripping of Dwarf Galaxies}

There are a number of indications that the group environment has an affect on dwarf galaxies.  In combination, dSph galaxies in the Local, Centaurus A, and Sculptor groups have mean distances of $0.23 \pm 0.20$ Mpc from the nearest large spiral galaxy, while transition-type and dIrr galaxies have mean distances of $0.54 \pm 0.31$ Mpc and $0.85 \pm 0.55$ Mpc, respectively \citep{2009AJ....138.1037C}.  These transition dwarfs are basically gas-rich dSphs and accordingly they fall on the luminosity-metallicity diagram near other dSphs \citep{1998ARA&A..36..435M}.  \citet{2009ApJ...696..385G} examine the \hi content of Local Group dwarf galaxies and find that within $\sim 270$ kpc, of either the Milky Way or Andromeda, all dwarfs (except more massive dwarf ellipticals) are undetected in \hi while most galaxies at larger radii have $10^5 - 10^8\ \msol$ of neutral gas.  The flat (compared to the field \himf) low mass slope of the \himf for the group environment \citep{2009MNRAS.400.1962K,2009AJ....138..295F, 2005ApJ...621..215S,2005nfcd.conf..351K, 2002A&A...382...43D, 2000ASPC..218..263V} is another indication that the neutral gas content of low mass galaxies is being altered.  

A galaxy moving through intergalactic gas will experience a drag force which, if strong enough, can strip gas from inside the galaxy potential.  The Gunn and Gott condition for a spiral galaxy balances the ram-pressure with the gravitational force per unit area in the plane of the galaxy from stars and gas \citep{1972ApJ...176....1G}, however, it does not account for the extra gravitational restoring force from the dark matter in the galaxy.  We use a slightly modified form \citep{2009ApJ...696..385G,2003AJ....125.1926G}, 
\begin{equation}
3v^2_{gal} n_{\mbox{\tiny IGM}}\sim \sigma_*^2 n_{gas} 
\end{equation}
where $n_{\mbox{\tiny IGM}}$ is the density of the intragroup medium, $v_{gal}$ is the velocity of the dwarf galaxy, $n_{gas}$ is the central \hi density of the dwarf galaxy, and $\sigma_*$ is the central stellar velocity dispersion in the dwarf.  When the left hand side of the equation becomes larger than the right hand side gas is removed from the galaxy.  It is assumed that the stripping is instantaneous and that the intergalactic gas density encountered by the moving galaxy remains constant.  The use of this criterion as a relatively good approximation is supported by numerical models of gas stripping in clusters \citep{2000ApJ...538..559M}.  

Using the IGM densities measured here we can explore whether ram-pressure stripping in groups is strong enough to remove significant reservoirs of neutral interstellar gas from dwarf galaxies, a key step in producing dSphs.  To model dwarf galaxies we choose the range of total mass, \hi mass, $\sigma_v$, and interstellar neutral gas density shown in Table \ref{tab:dwarfs}.  We compute the combination of galaxy velocity and IGM density necessary to instantaneously strip the neutral gas from a given dwarf galaxy using Equation 3 and show these curves in Figure \ref{fig:dwarfs} for the three dwarfs differentiating them by \hi mass.  The density and velocity characteristics of four groups from this paper are plotted to show which will have the necessary conditions to strip \hi from the dwarfs.  Comparing these stripping curves for generic dwarf galaxies to the IGM densities and average velocities of groups in this paper, we see that the dwarfs with $M_{HI}=10^5\ \msol$ ($M_{tot}=10^6\ \msol$) and $M_{HI}=10^6\ \msol$ ($M_{tot}=10^7\ \msol$) are likely to be stripped in all four groups shown. There are many additional factors that could increase the effectiveness of ram-pressure stripping including feedback from star-formation, the cosmic ionizing background, the orbital path of the galaxy, and a fluctuating density distribution.

The densities we derive here suggest that ram-pressure may be a critically important process affecting the gas content of dwarf galaxies.  The question of whether these dwarfs would be stripped in the Local or nearby groups with confirmed morphology-density relations is slightly more complicated.  The Local Group is an order of magnitude less massive ($\sim 1-2 \times 10^{12}\ \msol$; \citet{2005AJ....129..178K}) and thus not directly comparable with the systems that we consider here.  We can consider Leo T, a dwarf galaxy in the Local Group which still has a significant amount of \hi ($\sim 4 \times 10^5\ \msol$) and is at a Galactocentric distance of 420 kpc.  Leo T is likely to be similar to the progenitors of the newly discovered Milky Way satellites \citep{2010AdAst2010E..21W} all of which are at radii $< 250$ kpc and have no detected \hi.  \citet{2009ApJ...696..385G} estimate that Leo T would have to experience a halo gas density of $0.6-2.3 \times 10^{-3} \cmc$ to entirely remove its current \hi content.  We consider ourselves to be probing intergalactic gas using the radio sources in this paper, however, for systems like the Local Group the dividing line between halo gas in individual galaxies and IGM gas is unclear.   That dwarf galaxies in the Local Group could experience densities in this range is not impossible considering the IGM densities that we report here and the crossing time ($\sim 2$ Gyr) for the Milky Way system.     

In order to produce dSphs the neutral gas must be removed and the distribution of stars must also change.  Tidal stirring caused by repeated shocking at the pericenter of the dwarf's orbit can dissipate angular momentum, transforming the rotationally supported stellar distribution into one that is pressure supported \citep{2006MNRAS.369.1021M,2011arXiv1104.4278M}.  In these simulations, tidal heating increases the effectiveness of ram-pressure stripping by expanding the dwarf galaxy mass distribution.  

Ram pressure stripping is likely not strong enough to remove large quantities of neutral gas from galaxies with total masses of $\sim 10^9 - 10^{10}\ \msol$ which is where the \himf for groups of galaxies begins to flatten out.  In this regime, it may be more likely that these larger galaxies experience strong tidal stripping through interactions with neighboring galaxies similar to the interaction between the Large and Small Magellanic Clouds and the Milky Way.  In the group environment the space velocities are similar to the internal rotational velocities for large galaxies which makes tidal interactions especially damaging.  \hi observations of galaxy groups show numerous examples of intergalactic \hi that appears to be tidally stripped or large galaxies which are \hi deficient for their morphological type \citep{1994Natur.372..530Y,2001A&A...377..812V,2009AJ....138..295F}.

\begin{figure}
\plotone{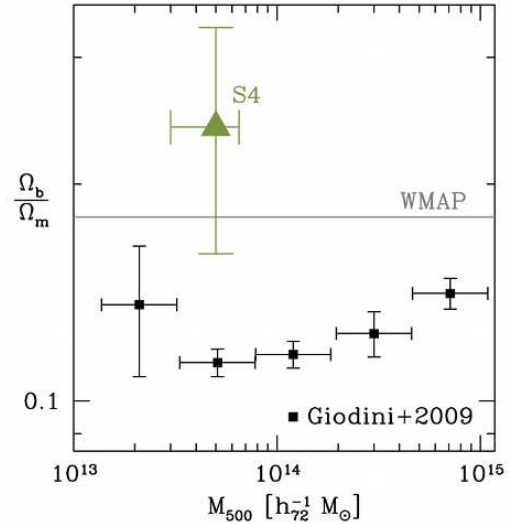}
\caption{A comparison of the baryon fraction for group S4 with binned data for groups and clusters from \citet{2009ApJ...703..982G}.  The mass for the S4 group is a dynamical mass derived from the velocity dispersion while the binned data are X-ray derived system masses.  For S4 the error bars reflect the range in dynamical mass (given the range in velocity dispersion) and the range in baryon fraction (given lower limits on the IGM gas mass and stellar mass in galaxies).  Note that the binned data are X-ray derived system masses and our S4 group mass is a dynamical mass.  We have not included a contribution from intragroup stars.  S4 is located at a radius of $700$ kpc from the group center and is probing gas with a density of $50\rho_{crit}$.  \textit{Galaxy groups contain significant reservoirs of baryons in their intragroup medium.} }\label{fig:missbar}
\end{figure}

\subsection{Strangulation}
The stripping of hot gas from the halos of galaxies preventing this gas from providing a fuel source for further star-formation is referred to as strangulation or starvation.  There are a handful of observations of X-ray tails from galaxies in a group environment \citep{2008ApJ...679.1162J,2007ApJ...664..804M,2006MNRAS.370..453R,2005ApJ...630..280M,2004ApJ...617..262S}.  Most simulations focus on ram-pressure stripping of disk and halo gas from galaxies in the cluster environment; there are few that consider low mass groups.  \citet{2008ApJ...672L.103K} perform a cosmological chemodynamical simulation of a $M \sim 3 \times 10^{11}\ \msol$ total mass galaxy in a group with $M \sim 8 \times 10^{12}\ \msol$ whose $\rho_{\tiny IGM}$ ranges from $2-5 \times 10^{-28} \gcmc$ ($2-5 \times 10^{-4} \cmc$, similar to the densities presented here).  They find that the star-formation rate in this galaxy decreases because most of the hot gas is stripped and cannot cool and form stars.  \citet{2009MNRAS.399.2221B} indicate that ram-pressure stripping of $> 50\%$ of the halo gas can occur in groups if $\rho_{\mbox{\tiny IGM}} \sim 7 \times 10^{-3} \cmc$ which is likely only the case in the very core of these systems. 

X-ray observations of galaxies in groups find that $80\%$ of the more massive galaxies ($L_K > L_*$) retain, to some extent, a hot gas halo although they appear to be more faint than field galaxies \citep{2008ApJ...679.1162J}.  They do not report the extent of these X-ray halos although the handful of galaxies which show X-ray tails or asymmetries have lengths that range from $6-50$ kpc.  \citet{2008MNRAS.383..593M} present an analytical stripping condition that describes what they see in their hydrodynamic simulations of stripping in clusters.  For a Milky Way sized galaxy ($M_{tot}\sim 10^{12}\ \msol$) moving with a velocity of $300 \kms$ through gas that is the same or ten times the density of its halo gas the halo will be stripped to a radius of $75$ kpc or $7.5$ kpc, respectively.  In the analytical model of \citet{2006ApJ...647..910H} the hot gas halo is easily stripped from galaxies in the group environment.  More observations are necessary to establish and compare the extent and brightness of hot halos in field and group galaxies.

\subsection{Baryon Content}

The baryon deficit in the local universe appears to scale with potential well depth such that the baryon content of massive clusters is nearly as expected while groups and individual galaxies are lacking \citep{2003ApJ...585L.117B, 2010ApJ...719..119D}.  There are recent indications that the possible reservoir of undetected gas in halos around galaxies is not sufficient to make up the balance \citep{2010ApJ...714..320A}.  

X-ray observations of intergalactic gas in clusters and more massive groups of galaxies are able to trace the radial density distribution of the hot IGM.  However, the current generation of X-ray telescopes are not able to detect emission from intergalactic gas in low mass galaxy groups or emission at large radii in more massive systems, likely because of the temperature of this gas.  In F08 we put an upper limit of $2 \times 10^6$ K on the temperature of the IGM at the location of bent-double radio source S1 using our density measurement and a non-detection in the X-ray data; that upper limit holds under the analysis in this paper.  

With this method we are able to probe the gas density at a single location in a given group but would still like to estimate the total mass in intergalactic gas.  In what follows we make an estimate of this mass and the baryon fraction of the S4 group.  We choose the S4 group for this exercise because its properties are most similar to the groups studied in \citet{2009ApJ...703..982G} and, in general, it's best to probe the baryon fraction at the largest possible radius.  The location of S4 at $R_{group}=700$ kpc allows us to probe the system at twice the radius possible for the S1 or S3 groups.  By assuming a uniform density within the radius of the bent-double radio source we put a lower limit on the mass in intergalactic gas of $M_{\mbox{\tiny IGM}}> 1 \times 10^{13}\ \msol$.  Using the velocity dispersion and its $68\%$ confidence intervals we estimate a range in dynamical mass for the S4 group of $M_{dyn} = 5_{-1}^{+2}\times 10^{13}\ \msol$.  We calculate a lower limit on the stellar mass of group galaxies ($M_* > 1.3 \times 10^{12}\ \msol$) by using the MPA/JHU value-added DR7 SDSS catalog of stellar masses (determined from photometry) as only the four brightest group members are present in the catalog. 
 
In Figure \ref{fig:missbar} we compare the baryon fraction estimated above for the S4 group with the COSMOS study of baryon mass fractions in groups and clusters with $z \leq 1$ \citep{2009ApJ...703..982G}.  They compute the baryon fraction at $R_{500}$ ($\gtrsim 500$ kpc for most groups) including contributions from intergalactic gas (from a fit to the X-ray gas mass fraction for 41 well-detected systems), galaxy stellar masses, and intergalactic stars.  Their total mass estimate for a system comes from an $L_X-M_{200}$ relation established by way of a weak lensing analysis.  The $\Lambda$CDM cosmological baryon fraction by mass is $0.18$ \citep{2011ApJS..192...18K} assuming a Hubble constant of $73 \kmsmpc$.  For source S4, located at a radius of 700 kpc this corresponds to a range in baryon fraction of $0.24^{+0.09}_{-0.08}$ which is consistent with what is expected cosmologically although we have not taken into account the errors on the IGM density.  
 
While we are unable to tightly constrain the baryon content of the intragroup medium in these systems, the gas densities that we measure here are valuable as complementary, model and temperature independent, measurements to X-ray and UV absorption line observations of the IGM in galaxy groups.  They indicate intergalactic gas does exist in galaxy groups, even at large radii, and can account for a large fraction of the baryons which are thought to be missing from these systems.

\section{Summary}

We use bent-double radio sources, under the assumption that their jets are bent by ram-pressure from their movement through intergalactic gas, to probe intragroup medium gas densities in galaxy groups.  This method provides a direct measurement of the intergalactic gas mass density that is complementary to UV absorption line and X-ray observations.  It allows us to probe intergalactic gas at large radii and in systems whose IGM is too cool to be detected by the current generation of X-ray telescopes.  

\renewcommand{\labelitemi}{$-$}

\begin{itemize}
\item We detect gas with significant densities $10^{-3}-10^{-4} \cmc$ in galaxy groups with velocity dispersions of a few hundred kilometers per second at radii ranging from $15-700$ kpc.\\
\item The combination of galaxy velocity and intragroup medium density in these systems is high enough to strip the neutral gas from dwarf galaxies with total masses of $10^{6-7}\ \msol$.  This stripping, in combination with the tidal interactions that are likely in this environment, can explain the existence of dwarf spheroidal galaxies.  Although we cannot make a direct connection between these systems and less massive groups, given the densities observed here it appears that ram-pressure stripping should also be effective in establishing the locally observed morphology-density relations for dwarf galaxies.
\item These observations indicate that intergalactic gas exists even in galaxy groups whose IGM remains undetected in X-rays, likely because the temperature of this gas places it below the sensitivity limit of current telescopes.  A rough estimate of the total baryonic mass in intergalactic gas is consistent with the missing baryons being located in the intragroup medium of galaxy groups.
\end{itemize}

\acknowledgements 

We thank the anonymous referee for comments which improved the paper.  We
acknowledge productive discussions with T. L. Irwin, J. Kormendy, S. Heinz, and
K.-V.  Tran.  This research has made use of the NASA/IPAC extragalactic database
(NED) which is operated by the Jet Propulsion Laboratory, Caltech, under
contract with the National Aeronautics and Space Administration.  We thank the
staff of the GMRT who have made these observations possible. GMRT is run by the
National Centre for Radio Astrophysics of the Tata Institute of Fundamental
Research.  The National Radio Astronomy Observatory is a facility of the
National Science Foundation operated under cooperative agreement by Associated
Universities, Inc.  Funding for the Sloan Digital Sky Survey (SDSS) and SDSS-II
has been provided by the Alfred P. Sloan Foundation, the Participating
Institutions, the National Science Foundation, the U.S. Department of Energy,
the National Aeronautics and Space Administration, the Japanese Monbukagakusho,
and the Max Planck Society, and the Higher Education Funding Council for
England. 


\end{document}